\documentclass[%
reprint,
superscriptaddress,
amsmath,amssymb,
]{revtex4-1}

\newcommand{\RomanNumeralCaps}[1]
 {\MakeUppercase{\romannumeral #1}}

\usepackage{graphicx}
\usepackage{dcolumn}
\usepackage{bm}
\usepackage{textcomp}
\usepackage{siunitx}

\begin{document}


\title{Nonlinear silicon waveguides generating broadband, spectrally engineered frequency combs spanning 2.0--8.5~\textmu{}m}


\author{Nima~Nader}
\email{nima.nader@nist.gov}
\affiliation{Applied Physics Division, National Institute of Standards and Technology, 325 Broadway, Boulder, Colorado 80305, USA}

\author{Abijith~Kowligy}
\thanks{These authors contributed equally to this work.}
\affiliation{Time and Frequency Division, National Institute of Standards and Technology, 325 Broadway, Boulder, Colorado 80305, USA}

\author{Jeff~Chiles}
\thanks{These authors contributed equally to this work.}
\affiliation{Applied Physics Division, National Institute of Standards and Technology, 325 Broadway, Boulder, Colorado 80305, USA}

\author{Eric~J.~Stanton} 
\affiliation{Applied Physics Division, National Institute of Standards and Technology, 325 Broadway, Boulder, Colorado 80305, USA}

\author{Henry~Timmers} 
\affiliation{Time and Frequency Division, National Institute of Standards and Technology, 325 Broadway, Boulder, Colorado 80305, USA}

\author{Alexander~J.~Lind} 
\affiliation{Time and Frequency Division, National Institute of Standards and Technology, 325 Broadway, Boulder, Colorado 80305, USA}
\affiliation{Department of Physics University of Colorado, 2000 Colorado Avenue, Boulder, Colorado 80309, USA}

\author{Flavio~C.~Cruz} 
\affiliation{Time and Frequency Division, National Institute of Standards and Technology, 325 Broadway, Boulder, Colorado 80305, USA}
\affiliation{Instituto de Fisica Gleb Wataghin, Universidade Estadual de Campinas, Campinas, SP, 13083-859, Brazil}

\author{Daniel~M.~B.~Lesko} 
\affiliation{Time and Frequency Division, National Institute of Standards and Technology, 325 Broadway, Boulder, Colorado 80305, USA}
\affiliation{Department of Physics University of Colorado, 2000 Colorado Avenue, Boulder, Colorado 80309, USA}

\author{Kimberly~A.~Briggman} 
\affiliation{Applied Physics Division, National Institute of Standards and Technology, 325 Broadway, Boulder, Colorado 80305, USA}

\author{Sae~Woo~Nam} 
\affiliation{Applied Physics Division, National Institute of Standards and Technology, 325 Broadway, Boulder, Colorado 80305, USA}

\author{Scott~A.~Diddams} 
\email{scott.diddams@nist.gov}
\affiliation{Time and Frequency Division, National Institute of Standards and Technology, 325 Broadway, Boulder, Colorado 80305, USA}
\affiliation{Department of Physics University of Colorado, 2000 Colorado Avenue, Boulder, Colorado 80309, USA}

\author{Richard~P.~Mirin}
\email{richard.mirin@nist.gov}
\affiliation{Applied Physics Division, National Institute of Standards and Technology, 325 Broadway, Boulder, Colorado 80305, USA}


\begin{abstract}
Nanophotonic waveguides with sub-wavelength mode confinement and engineered dispersion profiles are an excellent platform for application-tailored nonlinear optical interactions at low pulse energies.  Here, we present fully air clad suspended-silicon waveguides for infrared frequency comb generation with optical bandwidth limited only by the silicon transparency. The achieved spectra are lithographically tailored to span  2.1  octaves  in  the  mid-infrared (2.0--8.5~\textmu{}m or 1170--5000~\si{\per\cm}) when pumped at 3.10~\textmu{}m with 100~pJ pulses. Novel fork-shaped couplers provide efficient input coupling with only 1.5~dB loss. The coherence, brightness, and the stability of the generated light are highlighted in a dual frequency comb setup in which individual comb-lines are resolved with 30~dB extinction ratio and 100~MHz spacing in the wavelength range of 4.8--8.5~\textmu{}m (2100--1170~\si{\per\cm}). These sources are used for broadband gas- and liquid-phase dual-comb spectroscopy with 100~MHz comb-line resolution. We achieve a peak spectral signal-to-noise ratio of 10~\si{\sqrt{Hz}} across a simultaneous bandwidth containing 112,200 comb-lines. These results provide a pathway to further integration with the developing high repetition rate frequency comb lasers for compact sensors with applications in chip-based chemical analysis and spectroscopy.
\end{abstract}

\pacs{Valid PACS appear here}
\maketitle


\section{Introduction}

Nanophotonic waveguides offer sub-wavelength mode confinement with effective modal area of $\sim$1~\textmu{}m\textsuperscript{2}. This enhances the nonlinear optical interactions significantly and enables efficient wavelength conversion at low pulse energies. Subsequently, this results in the reduction of size, complexity, and power consumption of nonlinear optical systems. Moreover, the recent development of compact mode-locked lasers \cite{SchibliChipComb} with repetition rates, $f$\textsubscript{rep}, on the order of 1--10~GHz has increased the demand for integrated on-chip nonlinear devices operating at low pulse energies. Photonic waveguides have been used for carrier-offset frequency, $f_{0}$, detection and self-referencing of frequency combs using picojoule-level pulses \cite{SiNSelfRef,LipsonSiNSelfRef,LipsonOctaveSCGSelfRef,GaetaKeller19}. 

Optical frequency combs are phase-stabilized mode-locked lasers with broadband spectra of discrete, narrow optical lines spaced by the $f$\textsubscript{rep} \cite{diddams2000,JonesDiddamsScience2000,diddamsHgIonPRL2001,CundiffRevModPhys}. These are excellent sources for precision metrology and spectroscopy applications \cite{picqueMirNatPhot2012} where octave spanning, high-coherence, mid- and longwave-infrared (IR) spectral coverage are desired \cite{sorokina_solid-state_2003,ebrahim-zadeh_mid-infrared_2008}. Such sources access the molecular ro-vibrational states with resonant absorption lines unique to each molecule within the mid-IR (3--5~\textmu{}m) \cite{EOSMirPPLN,YcasMirNatPhot2018,KonstantinOPO2018,AlexArxiv} and longwave-IR molecular fingerprinting region (6--20~\textmu{}m) \cite{TimmersOptica,abijithEOS}. These advantages enable coherent probing of multiple ro-vibrational transitions in gas and condensed phases with an unprecedented frequency accuracy. Such lasers, implemented in the form of low-cost on-chip platforms, have applications as portable and mobile sensors for in-lab analysis, and fieldable trace chemical monitoring \cite{SinclairVan}.

In this paper, we utilize fully air clad suspended-Si nanophotonic waveguides \cite{ChilesASOP,KouSuspendedSi} for longwave-IR frequency comb generation. Si benefits from a broad transparency window of 1.1--8.5~\textmu{}m and a high nonlinear index, $n_2$, ($\sim100\times$ that of silica) \cite{n2SilicaOL1994,nonlinearParamSi2013}. The mature Si infrastructure enables reliable fabrication of versatile, low-loss waveguides suitable for on-chip efficient $\chi^{(3)}$ nonlinear processes. When pumped with 100~pJ, 85~fs pulses at 3.10~\textmu{}m, we generate IR spectra with 1--2~mW average powers and 2.1-octave bandwidth, spanning 2.0--8.5~\textmu{}m (1170--5000~\si{\per\cm}). Using the suspended-Si waveguides, we construct a dual-comb spectroscopy (DCS) setup for gas and liquid-phase sensing. Different waveguide widths enable measuring $\sim$\si{\num{2.7e5}} comb-lines that span the spectral range of 4.8--8.5~\textmu{}m (2100--1170~\si{\per\cm}), providing 100 MHz (0.0033~\si{\per\cm}) spectral resolution. The broadest measured simultaneous bandwidth spans 4.8--8.0~\textmu{}m with \si{\num{2.5e5}} comb-lines. We study atmospheric water absorption with the 100~MHz comb-line resolution and broadband absorption spectra of liquid-phase methanol and isopropanol. The smooth waveguide spectrum enables broadband baseline correction to realize excellent agreement with existing Fourier transform infrared (FTIR) measurements. 

This work improves the current state-of-the-art on-chip mid-IR generation technologies \cite{GaetaMirSiSCG2014,picqueMirSCG2015,EggletonSOS,SiuRes3umGaeta,SiuResDCSGaeta,NimaSOS,HicksteinAlN,AbijithPPLNwg,KippenbergMirSiN,SiGeOptica2018,Kippenberg4um} to realize picojoule-scale pulse energy and extends the spectral bandwidth to longwave-IR region. This is the first demonstration, to the best of our knowledge, of an on-chip IR frequency comb with optical bandwidth spanning 2.0--8.5~\textmu{}m, milliwatt scale average power, and a measured DCS figure-of-merit (FOM) equivalent to state-of-the-art mid-IR systems.

\section{Waveguide Design and Fabrication}

Si-photonics is predominantly focused on telecom applications based on the Si-on-insulator (SOI) material platform. It is, however, challenging to realize high performance nonlinear devices, mainly due to two-photon absorption (TPA) when waveguides are pumped below the TPA cutoff of 2.2~\textmu{}m. In addition, the SiO\textsubscript{2}-cladding has absorption at wavelengths $>$3.5~\textmu{}m, limiting the utility of this platform for mid- and longwave-IR applications \cite{GaetaMirSiSCG2014,picqueMirSCG2015,SiuRes3umGaeta}. A number of alternative approaches have been investigated based on modified cladding materials to reduce IR absorption in conjunction with longer-wavelength pump sources to eliminate the TPA. In particular, Si-on-sapphire waveguides pumped above the TPA cutoff \cite{LoncarSOS2013,EggletonSOS,NimaSOS} have shown promising results for mid-IR applications, but with bandwidths limited to 6~\textmu{}m, due to the sapphire absorption. Recently, SiGe-on-Si \cite{SiGeOptica2018} waveguides have been introduced as a new platform for spectral broadening reaching 8.5~\textmu{}m. The Si\textsubscript{x}Ge\textsubscript{y}/Si interface with 0.16 index contrast, however, limits the geometrical dispersion engineering.
 
\begin{figure}[t!]
\centering
 \includegraphics[width=1\linewidth]{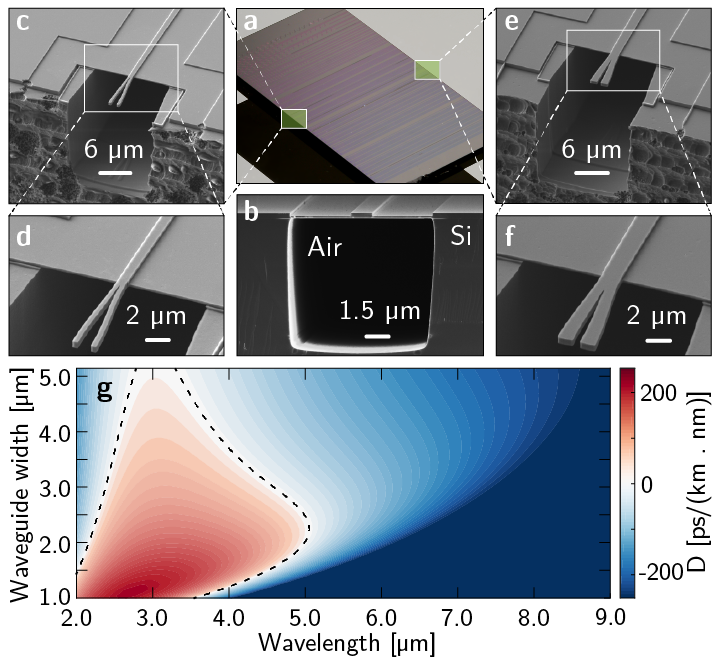}
 \caption{\label{FigWG} Fabricated suspended-Si waveguides. (a) Image of the fabricated suspended-Si chip with scanning electron micrograph (SEM) of (b) the suspended waveguide cross section, (c, d) The input and (e, f) the output fork-shaped couplers, respectively. (g) The designed GVD of the waveguides plotted as a function of the wavelength for different waveguide widths. The geometrical dispersion of the waveguides can be tailored to provide GVD zero-crossing at different wavelengths (dashed line), providing the phase matching for dispersive wave generation.}
\end{figure}

We design and fabricate suspended-Si waveguides based on 700~nm thick fusion-bonded Si membranes \cite{ChilesASOP} for supercontinuum generation (Fig.~\ref{FigWG}a, b). Removing the absorptive cladding eliminates the need for large cross-section waveguides to achieve low propagation losses in the IR \cite{SiLowLossLipson} and enables accessing the full transparency of Si. Moreover, this platform enables group-velocity-dispersion (GVD) engineering of the waveguides to realize application-tailored spectra through coherent dispersive wave generation. The bonded Si membrane is provided by a SOI wafer and the air trenches underneath the waveguides (Fig.~\ref{FigWG}b) are etched in a blank Si wafer prior to bonding (see supplement 1 for the fabrication details). The dimensions of the trenches are designed to avoid leakage losses of the generated longwave-IR light. The waveguides are formed by partial etching of the Si-membrane with etch depth of 390~nm, leaving a slab thickness of 310~nm to achieve the desired dispersion profiles for our nonlinear processes. 

We implement floating fork-shaped couplers for efficient input (Fig.~\ref{FigWG}c,~\ref{FigWG}d) and output (Fig.~\ref{FigWG}e,~\ref{FigWG}f) coupling between the waveguide chip and the free space mode. These are designed with two symmetrical arms to couple the free space mode to two points at the coupler tips. Such geometry enables fast compression of the optical mode into the waveguide, achieving high coupling efficiency and adiabatic operation in much shorter length-scales, compared to the conventional inverse tapers. Moreover, the floating structure enables controlled waveguide mode expansion in both horizontal and vertical directions. The widths of the coupler tips, along with their center-to-center gap can be designed to set the mode-field-diameter of the expanded mode for optimized mode-matching to free space. The compactness of the couplers make them highly desirable for suspended waveguide platforms, where mechanical stability places strong constraints on the dimensions of edge couplers.

We form the couplers by full etching the 700~nm thick suspended-Si membrane. To reduce scattering losses, a B\'{e}zier-type curvature is used to define the shape of the arms, while the widths of both arms are tapered from the tip to the point where they merge. The widths of the input and output taper tips are chosen as 440~nm, and 1.4~\textmu{}m and they taper to 440~nm and 900~nm in a floating length of 10~\textmu{}m, preserving the adiabatic operation of the couplers. The center-to-center gaps are also designed as 1.86~\textmu{}m and 2.2~\textmu{}m for input and output couplers, respectively. We measure the best input coupling efficiency of 1.50$\pm$0.13~dB/coupler at the 3.10~\textmu{}m pump wavelength and design the output couplers for broadband operation in 6.0--8.5~\textmu{}m range with 3~dB efficiency.

The high core-cladding index contrast of 3.4:1.0 in suspended-Si waveguides enables versatile geometrical GVD engineering with flat anomalous dispersion at the pump wavelength. The calculated GVD of the waveguides is presented in Fig.~\ref{FigWG}g as a function of wavelength for different waveguide widths. The plotted range of widths have anomalous dispersion at the pump wavelength, making the waveguides suitable for soliton fission and broadband coherent supercontinuum generation. Moreover, the long wavelength zero-dispersion wavelength (ZDW) can be tailored from 3.5~\textmu{}m to 5.0~\textmu{}m by varying the waveguide width, providing the phase matching condition for lithographically engineered dispersive wave generation in the longwave-IR.
 
\begin{figure}[t!]
\centering
 \includegraphics[width=1\linewidth]{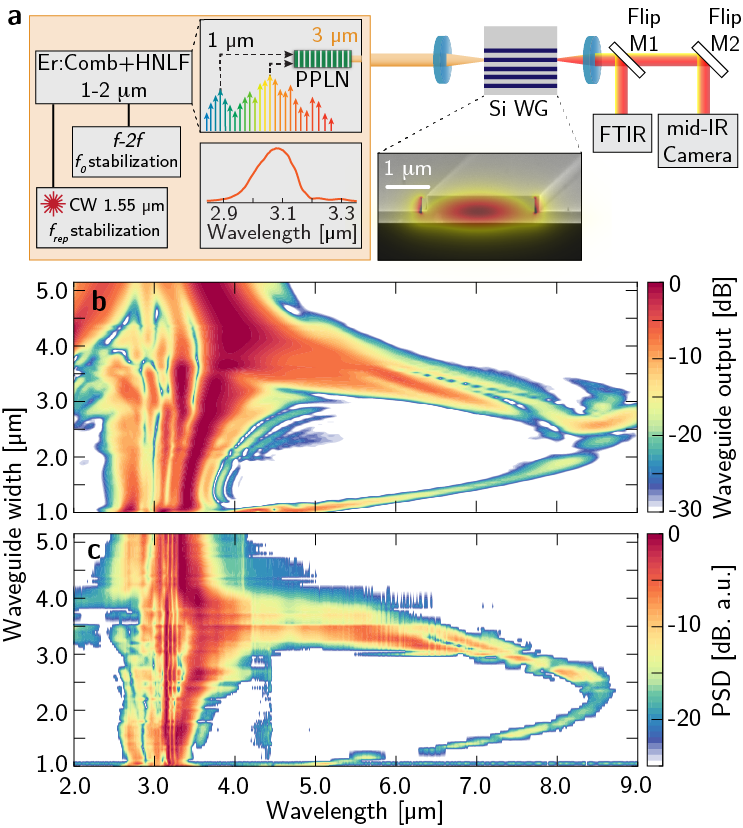}
 \caption{\label{FigSCG}Supercontinuum generation in the suspended-Si waveguides. (a) The schematic of the experimental setup for pumping the Si waveguides. The 3.10~\textmu{}m pump source is based on a 1550~nm Er:comb with $f_{0}$ and $f$\textsubscript{rep} locking and DFG in a PPLN crystal. HNLF: highly-nonlinear fiber, CW: Continuous wave, Si WG: silicon waveguide, M1 and M2: gold mirrors. A zoomed-in, cross sectional SEM of a suspended-Si waveguide overlaid with a simulation of the 8~\textmu{}m mode is also presented. The (b) calculated and (c) measured supercontinuum spectra of the waveguides.}
\end{figure}

For supercontinuum generation (SCG) experiments, we use a mid-IR laser with 85~fs pulses centered at 3.10~\textmu{}m operating with 100~MHz repetition rate (pump spectrum in Fig.~\ref{FigSCG}a). This source is based on a 1550~nm Er:doped oscillator and difference-frequency-generation (DFG) in a periodically-poled lithium niobate (PPLN) crystal \cite{Flavio3um}. We couple the free space beam of the 3.10~\textmu{}m pump laser to the TE\textsubscript{0} mode of the waveguides using a mid-IR Ge\textsubscript{28}Sb\textsubscript{12}Se\textsubscript{60} aspheric lens with numerical aperture of 0.56. The output is collected using a 0.82 numerical aperture lens and monitored with an InSb camera to optimize the alignment. The output lens is aligned to maximize the coupling for the longwave section of the spectrum at wavelengths $>$5.0~\textmu{}m and the collected supercontinuum spectra are recorded with an FTIR (Fig.~\ref{FigSCG}a). 

Pumping the waveguides at 3.10~\textmu{}m avoids TPA, and the nonlinear FOM increases by a factor of 4 compared with the TPA-limited value. This is a metric to calculate the trade-off between the nonlinearity of the medium, i.e., waveguides, and the nonlinear absorption. This parameter is defined as $n_{2}/\lambda\beta\textsubscript{TPA}$ and $n_{2}/\lambda I\beta\textsubscript{3PA}$ for TPA and three-photon-absorption (3PA), respectively \cite{SiFOM2011}. Here $\lambda$ is the pump wavelength and $I$ is the light intensity inside the waveguide. The TPA and 3PA nonlinear absorption coefficients are $\beta\textsubscript{TPA} = \num{0.65}~\si{cm/GW}$ \cite{Si2paCoeff} and $\beta\textsubscript{3PA} = \num{1.75e-3}~\si{cm^{3}/GW^{2}}$ \cite{nonlinearParamSi2013} at 1550~nm and the pump wavelength, respectively. The value \num{2.5e-5}~\si{cm^{2}/GW} is used for the nonlinear index, $n_{2}$ \cite{nonlinearParamSi2013}. 

\begin{figure*}[t!]
\centering
 \includegraphics[width=0.9\textwidth]{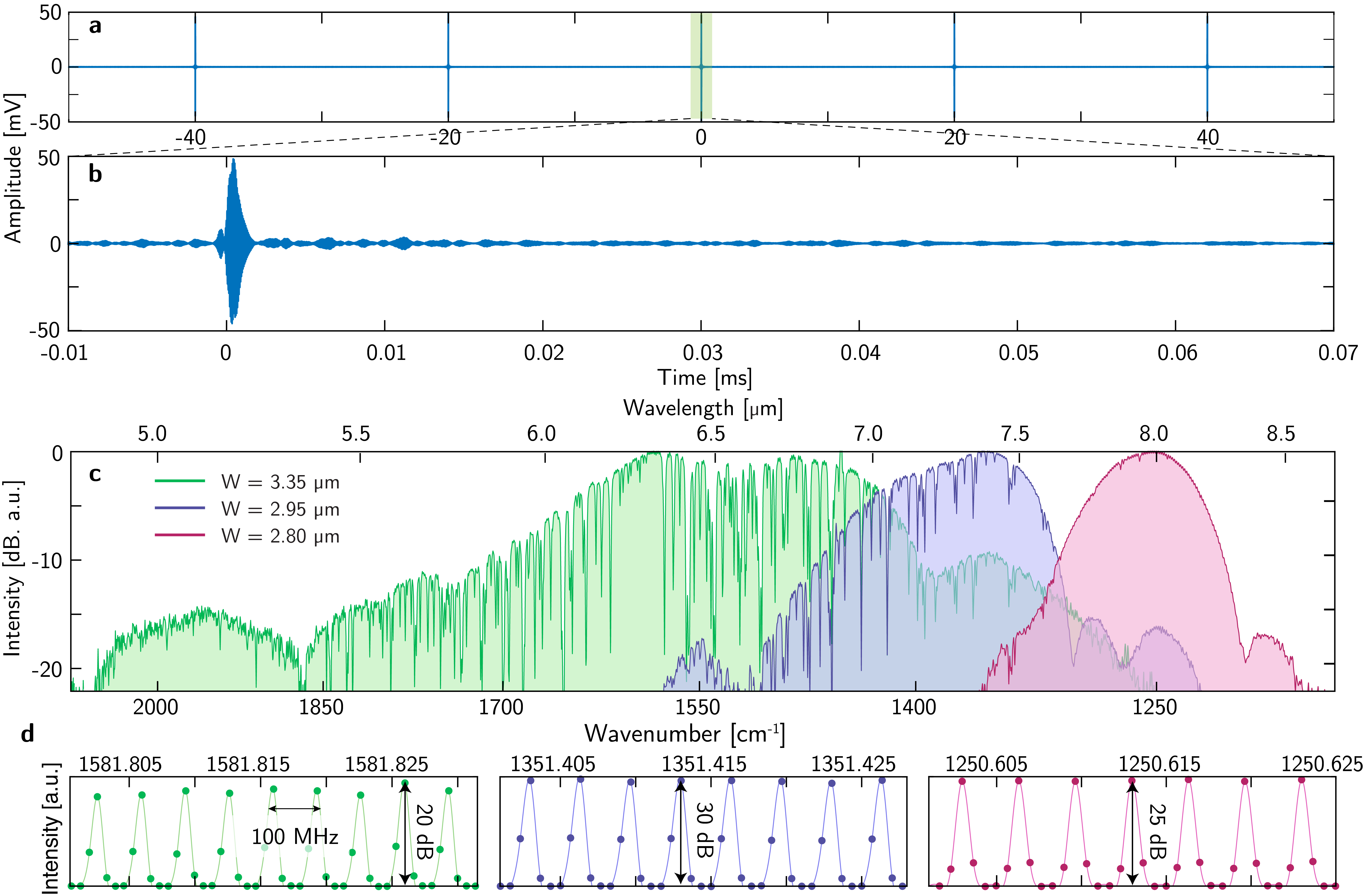}
 \caption{\label{FigCombline}Longwave-IR dual-comb interferograms and spectra. (a) The time-domain multi-heterodyne interferogram of the dual-comb system for the waveguide width of W = 2.95~\textmu{}m. A sequence of 5 interferograms is shown with time separation of 1/$\delta f$\textsubscript{rep} = 20~ms. (b) The zoomed-in view of one of the interferograms, emphasizing the center-burst and the trailing molecular free-induction decay oscillations of the atmospheric water absorption at the LWIR. (c) The frequency-domain dual-comb spectrum is calculated by taking an FFT of a single, averaged time-domain interferogram. Generated spectra for three waveguide widths of 2.80~\textmu{}m, 2.95~\textmu{}m, and 3.35~\textmu{}m are presented highlighting the spectral coverage from 4.8~\textmu{}m to 8.50~\textmu{}m. (d) A 0.025~\si{\per\cm} window of the dual-comb spectra in (c), presenting the 100~MHz spacing comb-lines resolved with 20--30~dB extinction ratio. This is calculated by taking an FFT of the five periodic time-domain intereferograms.}
\end{figure*}

We present the calculated and measured supercontinuum spectra of different waveguide widths with 100~pJ pulse energy coupled into the waveguide in Fig.~\ref{FigSCG}b and Fig.~\ref{FigSCG}c, respectively. The measured output optical powers of the waveguides range 1--2~mW (average power), depending on the waveguide widths. The theoretical supercontinuum is calculated by solving the generalized nonlinear Schr\"{o}dinger equation (gNLSE) \cite{AgrawalNLSE2007, NimaSOS}. The measured spectra are in excellent agreement with the simulations, enabling full control over the geometrical dispersion design parameters. For the waveguide widths of 1.0--3.0~\textmu{}m, the supercontinuum contains a dispersive wave that is lithographically tailored from 5.0--8.5~\textmu{}m, as predicted by the engineered long wavelength ZDW (Fig.~\ref{FigWG}g). Note that the narrowing of the longwave-IR dispersive wave bandwidth, above 8.0~\textmu{}m, is consistent with the absorption by Si phonon modes, not represented in our model. For wider waveguides (3.0--4.0~\textmu{}m widths) the broad and flat anomalous dispersion profiles with near-zero values result in broad supercontinuum generation covering more than an octave. The broadest bandwidth is measured for the waveguide width of 3.10~\textmu{}m and it spans 2.0--7.5~\textmu{}m. This broadband spectrum of comb-lines is suitable for spectroscopic study of a wide range of gas, liquid and solid phase samples.

\section{Dual-comb spectroscopy}

Frequency comb lasers have been extensively utilized for spectroscopy applications in the infrared. This includes complementing existing Fourier-transform spectrometers \cite{picqueCombFTIR,SpaunNat2016,BjorkDynamics,JYeC60} as well as the development of alternative spectroscopy schemes such as direct frequency comb spectroscopy (DFCS) with a highly dispersive etalon \cite{diddamsVipa2007,diddamsVipaOL2012,JYe3umVIPACombTransients,JYeMirVipa}, electro-optic sampling of the electric field \cite{ASOPSBartelsEOS,EOSMirPPLN,abijithEOS}, and dual-comb spectroscopy (DCS) \cite{HolzwarthLwirOL2004,IanOpticaDCS,OpgapLwirDCSDerryck,TimmersOptica,PicqueDCSreview2019}. 

DCS is an implementation of Fourier transform spectroscopy in which the interference of two frequency combs with slightly different repetition rates, $f$\textsubscript{rep} and $f$\textsubscript{rep}+$\delta f$\textsubscript{rep}, maps the optical spectrum to the radio frequency domain \cite{IanOpticaDCS}. Depending on the configuration, one or both combs pass through a sample and then interfere on a photodetector. In the time domain, an interference pattern is recorded whenever the pulses of the two combs overlap in time, generating a repetitive interferogram with period of 1/$\delta f$\textsubscript{rep}. The Fourier transform of this signal results in the optical spectrum of the dual-comb system with the imprinted sample absorption.

We use our waveguides in a dual-comb setup to demonstrate the coherence of the waveguide-generated light for DCS applications. The second comb in our DCS experiment is based on an Er:doped oscillator and intrapulse DFG in an orientation-patterned GaP (OP-GaP) crystal \cite{TimmersOptica}, generating $\sim$300~\textmu{}W of optical power in the wavelength range of 4--17~\textmu{}m. To better match the spectral bands, we add a short pass filter (MgF\textsubscript{2} window) to the beam path of the OP-GaP comb, limiting its long wavelength bandwidth to \textless9~\textmu{}m. In addition, we also place a 4.5~\textmu{}m long pass filter in the combined beam path of the two combs. The addition of these filters limits the spectral bandwidth of the dual-comb system to 4.5--8.5~\textmu{}m. The seed 1550~nm light of both combs are self-referenced for $f_{0}$ stabilization using conventional \textit{f}-to-2\textit{f} interferometers. In addition, using a 1550~nm cavity-stabilized continuous-wave laser, we stabilize repetition rates (Fig.~\ref{FigSCG}a) at the $\sim$100~nHz level (at 1~s) with $\delta f_{rep}$ = 50~Hz \cite{TimmersOptica}. This results in the dual-comb interferogram periodicity of 20~ms. We present a sequence of five interferograms in Fig. \ref{FigCombline}a, measured in a 100~ms acquisition time window using the output of the 2.95~\textmu{}m wide waveguide. The low amplitude noise and the high mutual coherence of the stabilized lasers enable us to achieve the estimated time-domain signal-to-noise ratio (SNR) of 1200 after averaging 16384 frames, corresponding to 27 minutes. Figure \ref{FigCombline}b presents a 80~\textmu{}s window of one of the interferograms. The large oscillation near t = 0~s is the center-burst representing the spectral envelop of the dual-comb system. The trailing oscillations are molecular free-induction-decay \cite{TimeDomainIan} signatures of the atmospheric water absorption centered at $\sim$6.25~\textmu{}m ($\sim$1600~\si{\per\cm}).

\begin{figure*}[t!]
\centering
 \includegraphics[width=1.0\textwidth]{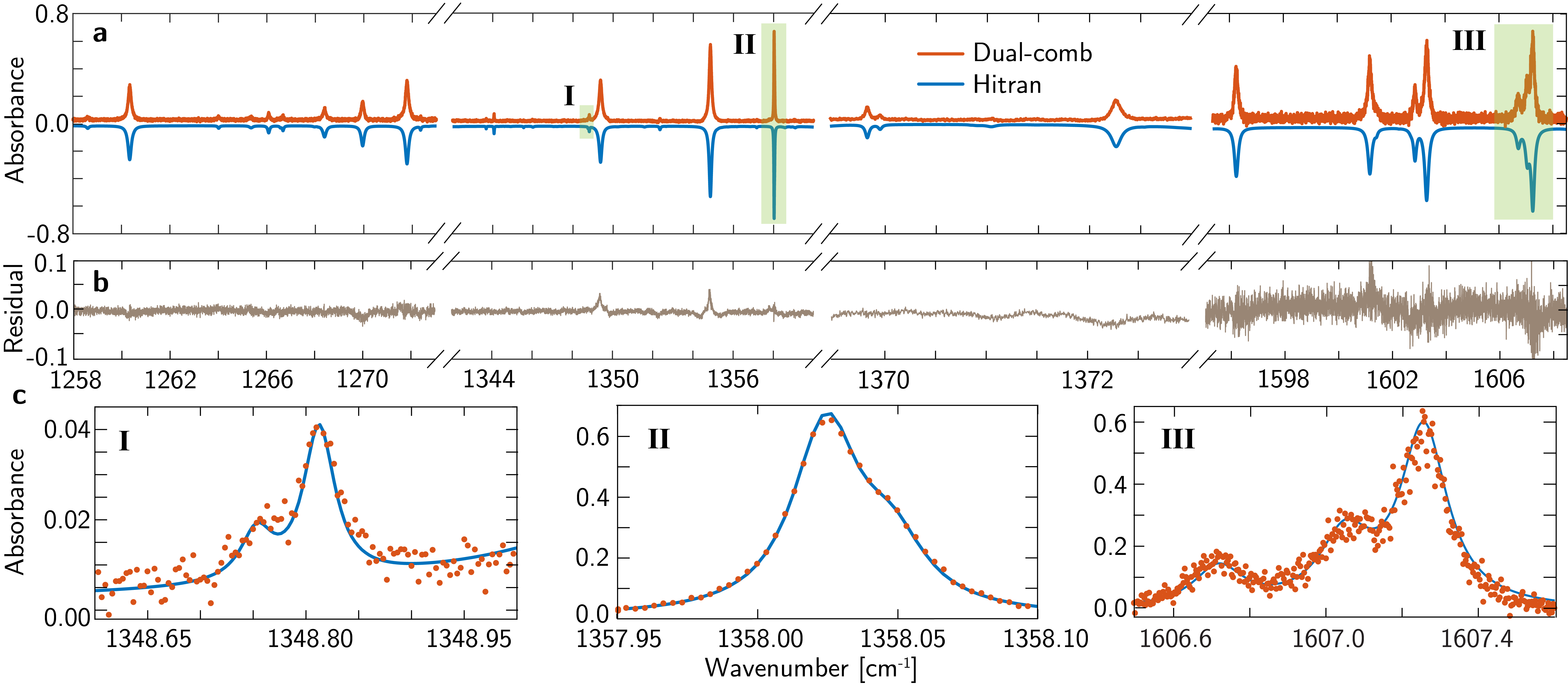}
 \caption{\label{FigDCSwater}Dual-comb measurement of the atmospheric water absorption. (a) Baseline-corrected DCS of the atmospheric water absorption with the 100~MHz (0.0033~\si{\per\cm}) comb-line spacing resolution in the 1250--1610~\si{\per\cm} range (6.2--8.0~\textmu{}m). The dual-comb data (red) is compared with the HITRAN database (blue, reflected about x-axis). Data is measured using 2.80~\textmu{}m, 2.95~\textmu{}m, and 3.35~\textmu{}m wide waveguides. The HITRAN is fitted to the DCS data by a Voigt fitting algorithm to estimate Lorentzian and Gaussian lineshapes. (b) Difference between the data and the fit (residuals), showing excellent agreement between DCS and HITRAN within 1.5\% uncertainty. The higher residuals at 1588--1610~\si{\per\cm} (5\% error) are due to the lower optical power from the dual-comb system in this frequency range. (c) The zoomed-in view of the highlighted regions in (a).}
\end{figure*}

We retrieve the optical spectrum by calculating the Fourier-transform of the time-domain interferogram. Using three different waveguide widths of 3.35~\textmu{}m, 2.95~\textmu{}m, and 2.80~\textmu{}m (Fig.~\ref{FigCombline}c) our dual-comb setup covers the infrared range of 4.8--8.5~\textmu{}m (2100--1170~\si{\per\cm}) with spectral resolution of 100~MHz (0.0033~\si{\per\cm}). The broadest simultaneous bandwidth is measured using the 3.35~\textmu{}m wide waveguide and it spans 4.8--8.0~\textmu{}m (1250--1200~\si{\per\cm}), containing \si{\num{2.7e5}} comb-lines. The highest spectral SNR is obtained using the 2.95~\textmu{}m wide waveguide, while the waveguide width of 3.35~\textmu{}m has the lowest measured SNR due to the broad bandwidth and the lower optical power per comb-line generated by this device. For the averaging time of $\tau$ = 330~s, we estimate the highest obtained SNR to be 180 over the 374~\si{\per\cm} spectral bandwidth. Normalized to a one second averaging time, we get $SNR$ = \num{10}~\si{\sqrt{Hz}} at 100~MHz comb-line resolution. Having a total of $M$~=~112,200 comb-lines, we calculate our DCS FOM as $M\times SNR~=$~\num{1.1e6}~\si{\sqrt{Hz}} \cite{NateDCSSensitivity}. This is similar to the previously reported value for a dual-comb setup with two identically designed infrared frequency combs based on intrapulse DFG in OP-GaP crystals \cite{TimmersOptica}, confirming the coherence of the nonlinear processes in the Si waveguides. Figure~\ref{FigCombline}d presents zoomed-in views of the three spectra in Fig.~\ref{FigCombline}c, emphasizing the resolved individual comb-lines with the 100~MHz spacing. Here the comb-lines are resolved with 20~dB, 25~dB, and 30~dB extinction ratios for waveguide widths of 3.35~\textmu{}m, 2.80~\textmu{}m, and 2.95~\textmu{}m, respectively.

We utilize the waveguide-based dual-comb setup to study atmospheric water vapor from 1200--1600~\si{\per\cm}. Such analysis enables monitoring of controlled lab environment through the study of atmospheric pressures and water concentrations, measured in terms of volume-mixing ratio. Figure~\ref{FigDCSwater}a presents the measured atmospheric water absorbance with 100~MHz comb-line resolution. Data, presented in red, is compared to the HITRAN database \cite{HITRAN2016}, presented in blue and reflected about x-axis. We define absorbance as $A = \mathrm{log_{10}}(I/I_{0})$, where $I_{0}$ and $I$ are the calculated spectral baseline (Supplement 1) and the measured spectrum, respectively. Our dual-comb setup has an atmospheric beam path of $\sim$2~\si{m}, resulting in many saturated absorbance peaks. Hence, we only present and analyze the absorption data in three unsaturated regions of 1258--1272~\si{\per\cm}, 1340--1373~\si{\per\cm}, and 1588--1608~\si{\per\cm}. The data in these ranges are measured using 2.80~\textmu{}m, 2.95~\textmu{}m, and 3.35~\textmu{}m wide waveguides, respectively.

We calculate the fit residuals as data minus the HITRAN model (Fig.~\ref{FigDCSwater}b). The calculated root-mean-square (RMS) of the residuals is 0.007 for the measurement ranges of 1258--1272~\si{\per\cm} and 1342--1373~\si{\per\cm}, emphasizing the excellent agreement between the DCS and HITRAN. This value increases for the 1588--1680~\si{\per\cm} range to 0.02 due to the lower optical power. Such excellent agreement between DCS measurements and the HITRAN database enables us to estimate the in-lab atmospheric pressure and volume-mixing ratio of the water content as 817$\pm$41~mbar and 0.016$\pm$0.002, respectively. These values agree well with the typical recorded values for the city of Boulder due to the area's higher elevation. A detailed explanation of this estimation is provided in Supplement 1.

To emphasize our high DCS figure-of-merit, excellent frequency resolution and low fitting uncertainty, we highlight three absorbance features in Fig.~\ref{FigDCSwater}a and present their zoomed-in windows in Fig.~\ref{FigDCSwater}c. The high DCS FOM enables us to detect absorbance features of 10\textsuperscript{-2} levels (Fig.~\ref{FigDCSwater}c, panel \RomanNumeralCaps{1}) which is similar to the levels achieved with FTIR spectrometers. Moreover, the fine comb-tooth spacing enables resolving narrow linewidth features such as presented in Fig.~\ref{FigDCSwater}c, panel \RomanNumeralCaps{2}, with FWHM~$\approx$~1.2~GHz (0.04~\si{\per\cm}). In a conventional FTIR spectrometer, a delay range of 3.0~m would be required to achieve the demonstrated 100~MHz spectral resolution. Despite the increased residual in 1588--1608~\si{\per\cm} range, the measurement still agrees well with HITRAN (Fig.~\ref{FigDCSwater}c, panel \RomanNumeralCaps{3}). We can estimate the in-lab atmospheric pressure and water concentration by only comparing the HITRAN model to this data set. In this case, the atmospheric pressure and water content are calculated within 1.4\% and 3\% of the values estimated using the data from all waveguide widths, respectively.

\begin{figure}[t!]
\centering
 \includegraphics[width=1.0\linewidth]{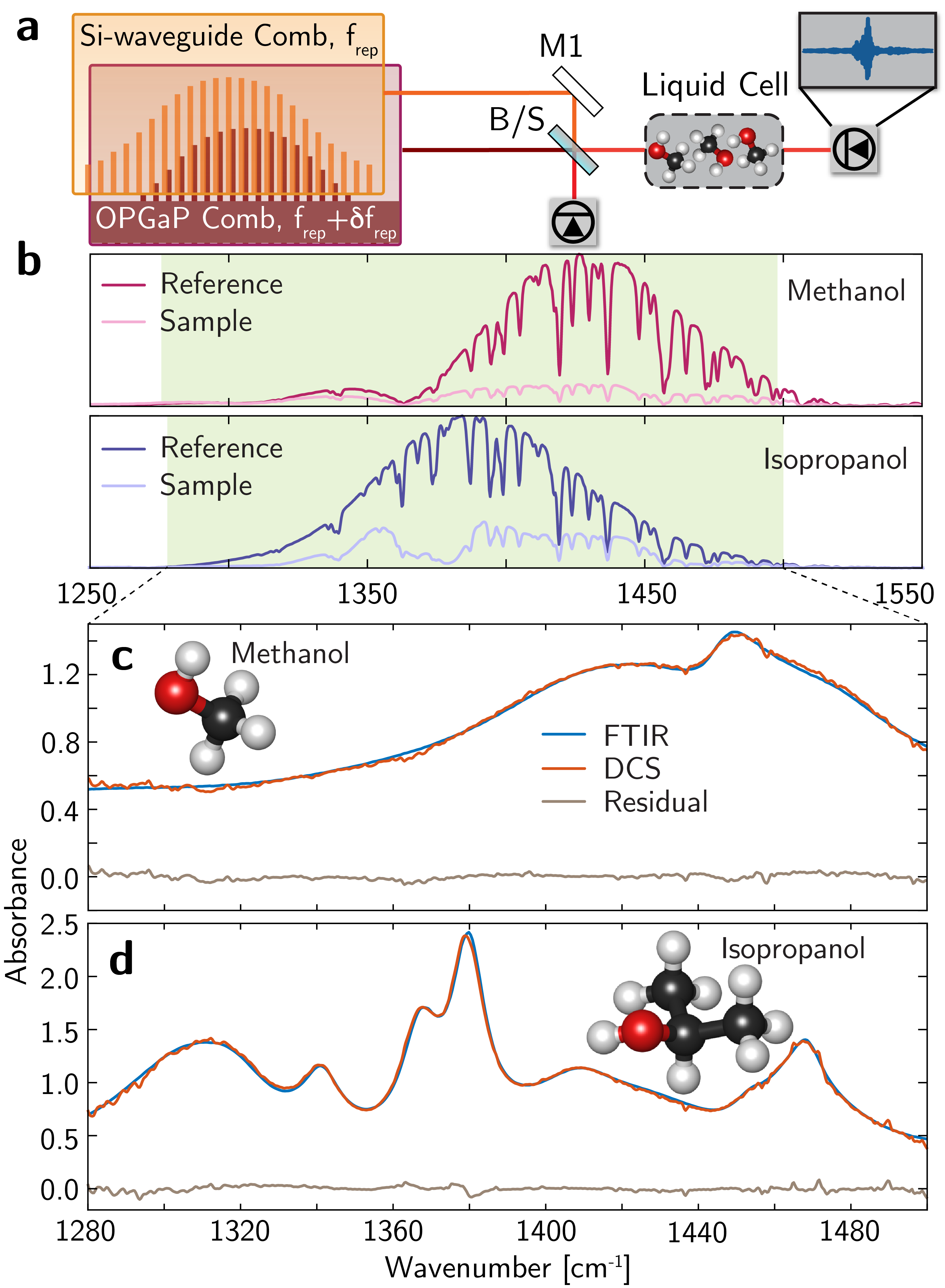}
 \caption{\label{FigDCSliquid}Broadband DCS of liquid methanol and isopropanol. (a) The schematic diagram of the dual-comb setup for simultaneous measurement of the sample and reference spectra. M1: gold mirror, B/S: mid-IR beamsplitter. (b) The dual-comb spectra of the reference and the sample for methanol (top panel) and isopropanol (bottom panel). The green shaded area is the region with sufficient SNR to calculate the absorbance of the liquid samples. The measured DCS absorbance of (c) methanol and (d) isopropanol are compared with FTIR measurements in the range of 1280--1500~\si{\per\cm}. The simultaneous reference and sample measurement along with the smooth waveguide spectrum enable broad bandwidth DCS of liquid samples.}
\end{figure}

We also measure the absorption spectrum of liquid phase alcohols, leveraging the broadband and smooth waveguide spectra. Characterization of broad absorbers require spectral envelope stability over a broadband region to enable proper baseline measurement and subtraction \cite{YcasMirNatPhot2018}. The stability of our supercontinuum-generated frequency comb enables coherent averaging and measurement of large absorbance values. Moreover, to eliminate the effects of long term, few minutes timescale, drifts we perform simultaneous measurement of the sample and reference spectra. In such a scheme, we use two liquid-nitrogen cooled mercury-cadmium-telluride (MCT) detectors at the two sides of the 50/50 beamsplitter that combines the infrared combs (Fig.~\ref{FigDCSliquid}a). The combined beams pass through a sample cell in one of the arms before being detected. In the other arm the beam is sent directly to the detector as the reference measurement. Care is taken to have equal beam paths between the two arms to minimize the residual atmospheric absorption features after the baseline subtraction.

We choose isopropanol and methanol for liquid-phase spectroscopy because they are widely used for scientific and industrial applications. For our DCS demonstration we use a 15~\textmu{}m thick liquid-sample cell to minimize the interaction length with the liquid and avoid saturated absorption. In liquid-phase spectroscopy the absorbance lines are broadened to form a continuous spectrum  and 1~\si{\per\cm} spectral resolution is sufficient to resolve the features. We perform the liquid-phase DCS experiments with a resolution of 0.67~\si{\per\cm}, achieved via temporal apodization of the dual-comb interferogram to a 100~\textmu{}s window. Fig.~\ref{FigDCSliquid}b presents the sample and reference spectra of methanol (top panel) and isopropanol (bottom panel), measured using different waveguide widths for each sample.

We compare our baseline subtracted (Supplement 1) DCS data to measurements performed using a commercial FTIR operating with 1~\si{\per\cm} resolution in Figs.~\ref{FigDCSliquid}c and Figs.~\ref{FigDCSliquid}d for methanol and isopropanol, respectively. The DCS data is in excellent agreement with the FTIR spectrum, yielding a RMS residual value of 0.02. We note that the calculated residual levels are only limited by the uncorrected atmospheric water absorption contaminating the DCS data. Such agreement enables accurate analysis of the DCS spectra to assign the measured absorbance features to different molecular vibrational transitions (Supplement 1). 

While demonstrated here for well-known alcohols, the bright waveguide-generated light and dual-comb measurement techniques we employ can be widely applied to other samples. The frequency range of 1200--1600~\si{\per\cm} enables access to C---H and O---H bending functional groups. This region of the infrared provides stronger integrated absorption intensities and lower ro-vibrational density of states, when compared with C---H stretching functional groups at 2000--3000~\si{\per\cm} (3--5~\textmu{}m).

\section{Conclusion and Summary}

We introduced suspended-Si waveguides as a versatile nonlinear photonic platform for spectral engineering of frequency combs across the mid- and longwave-IR. Our waveguide-generated comb light covers the optical bandwidth of 2.0--8.5~\textmu{}m with milliwatt-scale average power enabling access to the molecular fingerprinting region. Here, we leverage careful dispersion engineering and mature fabrication to demonstrate efficient photonic-chip based spectral engineering across a 115 THz bandwidth with a 100~pJ pump pulse energy. To the best of our knowledge, this is the first demonstration of nanophotonic-based frequency combs with such broadband spectra and milliwatt-scale average powers.

To demonstrate the coherence of the waveguide-generated light and its utility for DCS applications, we utilized our nonlinear devices in a dual-comb setup. We have achieved a DCS FOM of \num{1.1e6}~\si{\sqrt{Hz}} which is competitive with current state-of-the-art systems operating in the same spectral region. The smoothness of the generated spectra and our fine comb-tooth spacing of 100~MHz (0.0033~\si{\per\cm}) enables probing of gas-phase narrow linewidth absorbers as well as condensed-phase samples with broad absorption features. The DCS data have excellent agreement with the standard databases and FTIR measurements within less than 5\% error. 

Moreover, using the 2.80~\textmu{}m wide waveguide as an example, $\sim$20\% of the optical power at 8~\textmu{}m extends into the air trenches underneath the waveguide core. Such an easy, on-chip access to the longwave-IR optical power presents the opportunity to integrate these devices with chip-based chemical delivery systems like microfluidic channels. We envision the integration of this platform with high repetition rate laser frequency combs and chemical delivery schemes for on-chip, parallel, multichannel \textit{in situ} chemical studies and reaction monitoring.

\paragraph*{\textbf{\textup{Funding:}}} This work is supported by NIST and the Defense Advanced Research Projects Agency (DARPA), Defense Sciences Office (DSO) under the SCOUT program.

\paragraph*{\textbf{\textup{Acknowledgements:}}} We thank Travis Autry, Fabrizio R. Giorgetta, Esther Baumann, Jeffrey Shainline, and David Carlson for useful discussions and inputs on the manuscript. F.C.C. acknowledges funding from Fapesp (Grant~\#~2018/26673-5). This is a contribution of NIST, an agency of the US government, not subject to copyright. Product disclaimer: Any mention of commercial products is for information only; it does not imply recommendation or endorsement by NIST.

\bibliography{References}

\end{document}



\title{Nonlinear silicon waveguides generating broadband, spectrally engineered frequency combs spanning 2.0--8.5~\textmu{}m: supplementary material}


\author{Nima~Nader}
\email{nima.nader@nist.gov}
\affiliation{Applied Physics Division, National Institute of Standards and Technology, 325 Broadway, Boulder, Colorado 80305, USA}

\author{Abijith~Kowligy}
\thanks{These authors contributed equally to this work.}
\affiliation{Time and Frequency Division, National Institute of Standards and Technology, 325 Broadway, Boulder, Colorado 80305, USA}

\author{Jeff~Chiles}
\thanks{These authors contributed equally to this work.}
\affiliation{Applied Physics Division, National Institute of Standards and Technology, 325 Broadway, Boulder, Colorado 80305, USA}

\author{Eric~J.~Stanton} 
\affiliation{Applied Physics Division, National Institute of Standards and Technology, 325 Broadway, Boulder, Colorado 80305, USA}

\author{Henry~Timmers} 
\affiliation{Time and Frequency Division, National Institute of Standards and Technology, 325 Broadway, Boulder, Colorado 80305, USA}

\author{Alexander~J.~Lind} 
\affiliation{Time and Frequency Division, National Institute of Standards and Technology, 325 Broadway, Boulder, Colorado 80305, USA}
\affiliation{Department of Physics University of Colorado, 2000 Colorado Avenue, Boulder, Colorado 80309, USA}

\author{Flavio~C.~Cruz} 
\affiliation{Time and Frequency Division, National Institute of Standards and Technology, 325 Broadway, Boulder, Colorado 80305, USA}
\affiliation{Instituto de Fisica Gleb Wataghin, Universidade Estadual de Campinas, Campinas, SP, 13083-859, Brazil}

\author{Daniel~M.~B.~Lesko} 
\affiliation{Time and Frequency Division, National Institute of Standards and Technology, 325 Broadway, Boulder, Colorado 80305, USA}
\affiliation{Department of Physics University of Colorado, 2000 Colorado Avenue, Boulder, Colorado 80309, USA}

\author{Kimberly~A.~Briggman} 
\affiliation{Applied Physics Division, National Institute of Standards and Technology, 325 Broadway, Boulder, Colorado 80305, USA}

\author{Sae~Woo~Nam} 
\affiliation{Applied Physics Division, National Institute of Standards and Technology, 325 Broadway, Boulder, Colorado 80305, USA}

\author{Scott~A.~Diddams} 
\email{scott.diddams@nist.gov}
\affiliation{Time and Frequency Division, National Institute of Standards and Technology, 325 Broadway, Boulder, Colorado 80305, USA}
\affiliation{Department of Physics University of Colorado, 2000 Colorado Avenue, Boulder, Colorado 80309, USA}

\author{Richard~P.~Mirin}
\email{richard.mirin@nist.gov}
\affiliation{Applied Physics Division, National Institute of Standards and Technology, 325 Broadway, Boulder, Colorado 80305, USA}


\begin{abstract}
This document provides supplementary information to ``Nonlinear silicon waveguides generating broadband, spectrally engineered frequency combs spanning 2.0--8.5~\textmu{}m.'' This document contains four sections. Section 1 explains the details of the suspended-Si waveguides fabrication. Section 2 provides the details of baseline fitting and subtraction to study dual-comb measured atmospheric water absorbance data. We also explain the detailed procedure for the comparison of the data with HITRAN database to estimate the experimental atmospheric pressure and water content. Section 3 provides the detailed process for background subtraction of the broadband liquid-phase dual-comb data. Section 4 goes through the detailed spectral peak assignments of the measured vibrational lines in DCS of isopropanol and methanol.
\end{abstract}

\pacs{Valid PACS appear here}
\maketitle


\section{Waveguide Fabrication}

\begin{figure}[t!]
\includegraphics[width=\linewidth]{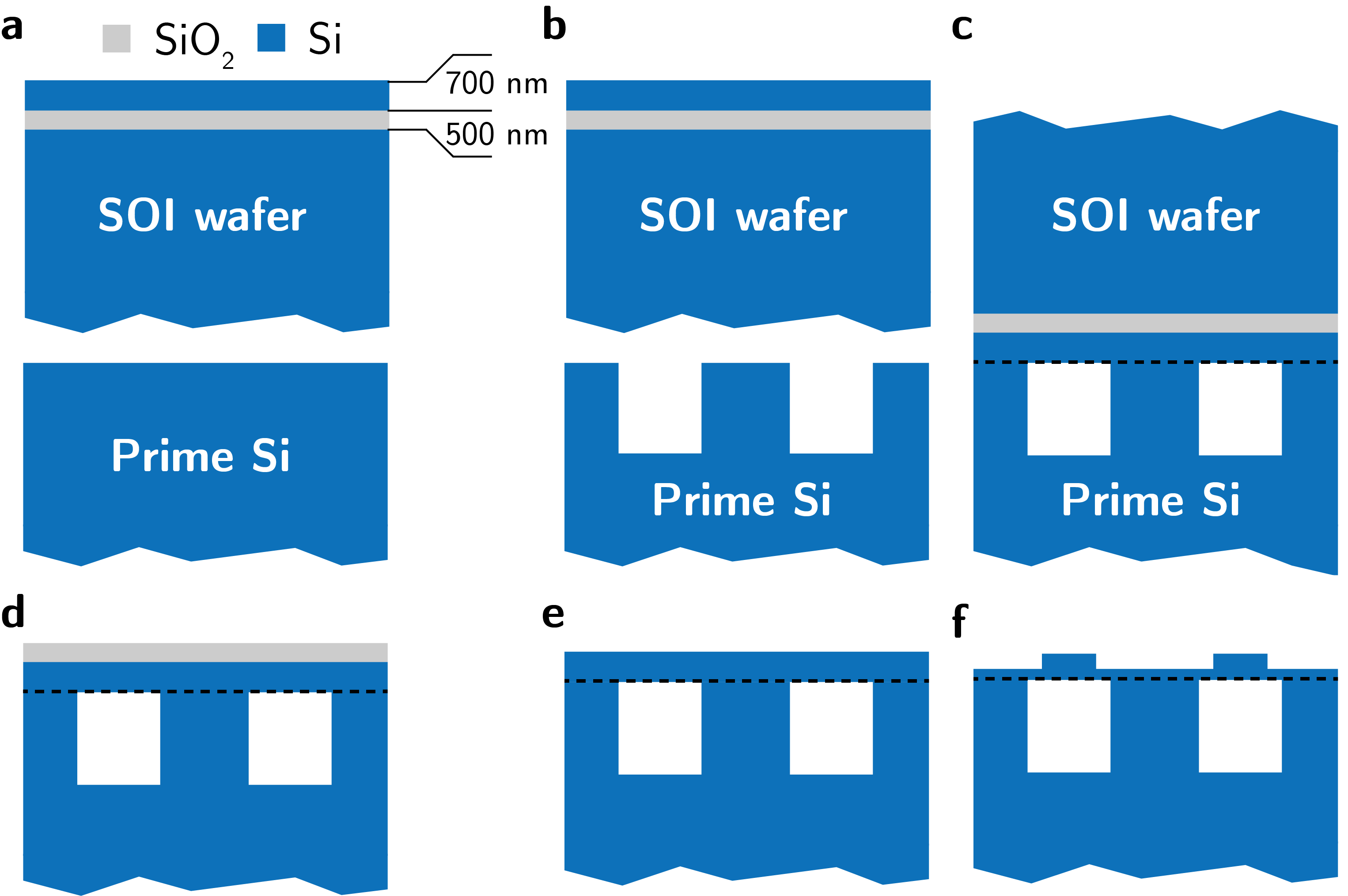}
\caption{\label{FigFab}Schematic diagram of the suspended-Si waveguide fabrication steps. (a) The prime-Si and the Si-on-insulator wafers used in the process. (b) RIE etching of the suspended waveguide trenches in the prime-Si wafer. (c) Wafer bonding. (d) Backside Si wafer removal with (e) BOE removal of the buried oxide layer. (f) Patterning and etching of the waveguide structures.}
\end{figure}

We fabricate suspended-Si waveguides following the method presented in \cite{ChilesASOP}. The process starts with two wafers (Fig.~\ref{FigFab}a), namely one prime-grade Si and a Si-on-Insulator (SOI). The SOI wafer has a 700~nm thick Si device layer on top and a 500~nm thick buried oxide underneath (Fig.~\ref{FigFab}a). Both wafers are solvent- and H\textsubscript{2}SO\textsubscript{4}:H\textsubscript{2}O\textsubscript{2}-cleaned prior to the processing. Suspended waveguide trenches are patterned in a positive tone photoresist by 405~nm lithography. These patterns are transferred to the prime-Si wafer with an etch depth of 15~\textmu{}m by reactive ion etching (RIE) with SF\textsubscript{6}:C\textsubscript{4}F\textsubscript{8} based chemistry (Fig.~\ref{FigFab}b). After this step, both wafers are prepared for wafer bonding as follows: The native oxide on both wafers is removed with a diluted, 5\%, buffered-oxide-etch (BOE), making both surfaces hydrophobic. Wafers are then transferred to an atmospheric plasma system for surface activation in a N\textsubscript{2}:H\textsubscript{2} plasma. When completed, this plasma activation makes the bonding interfaces hydrophilic. Wafers are then soaked in water to saturate the activated surfaces with --OH groups. After drying, wafers are placed in contact in an atmospheric wafer bonder to initiate the temporary bond (Fig.~\ref{FigFab}c), assisted by Van der Waals forces between the hydroxyl groups. To make the bond permanent, the wafer stack is annealed at 300~C for 6~hours in a N\textsubscript{2}-purged furnace. The annealing process replaces the OH--OH bonds with Si--O--Si fusion bonds and the released gases exit the bond interface through the suspended waveguide trenches. During this step, trenches act as gas release channels to minimize bond voids, and therefore increasing the fabrication yield significantly. We continue the processing by removing the backside Si wafer (Fig.~\ref{FigFab}d) and the buried oxide layer (Fig.~\ref{FigFab}e) with RIE etching and diluted BOE, respectively.

A 200~nm thick SiO\textsubscript{2} hard mask is deposited on the Si membrane perior to patterning and etching of the waveguides. A positive tone electron-beam (e-beam) lithography resist is spin-coated with the thickness of $\sim$~400~nm on the hard mask and waveguide structures are patterned using an e-beam writer. The patterns are RIE etched in the SiO\textsubscript{2} hard mask using a CHF\textsubscript{3} chemistry. Following the hard mask etch, e-beam resist is solvent-removed and the waveguide structures are transferred to the suspended-Si membrane using a HBr-chemistry RIE etch (Fig.~\ref{FigFab}f). The waveguide patterning ends with hard mask removal with 5\% BOE. The same step is repeated to pattern and etch the floating fork-shaped couplers but with minor differences. Namely, a thicker hard mask ($\sim$~300~nm) and  a thicker e-beam resist ($\sim$~780~nm) are used to support the full etch of the 700~nm thick Si membrane to achieve floating structures. The fabrication process ends with a die release step using a chemical dicing recipe based on deep-RIE etching of Si.

\section{Atmospheric water absorption}

\begin{figure}[t!]
\includegraphics[width=\linewidth]{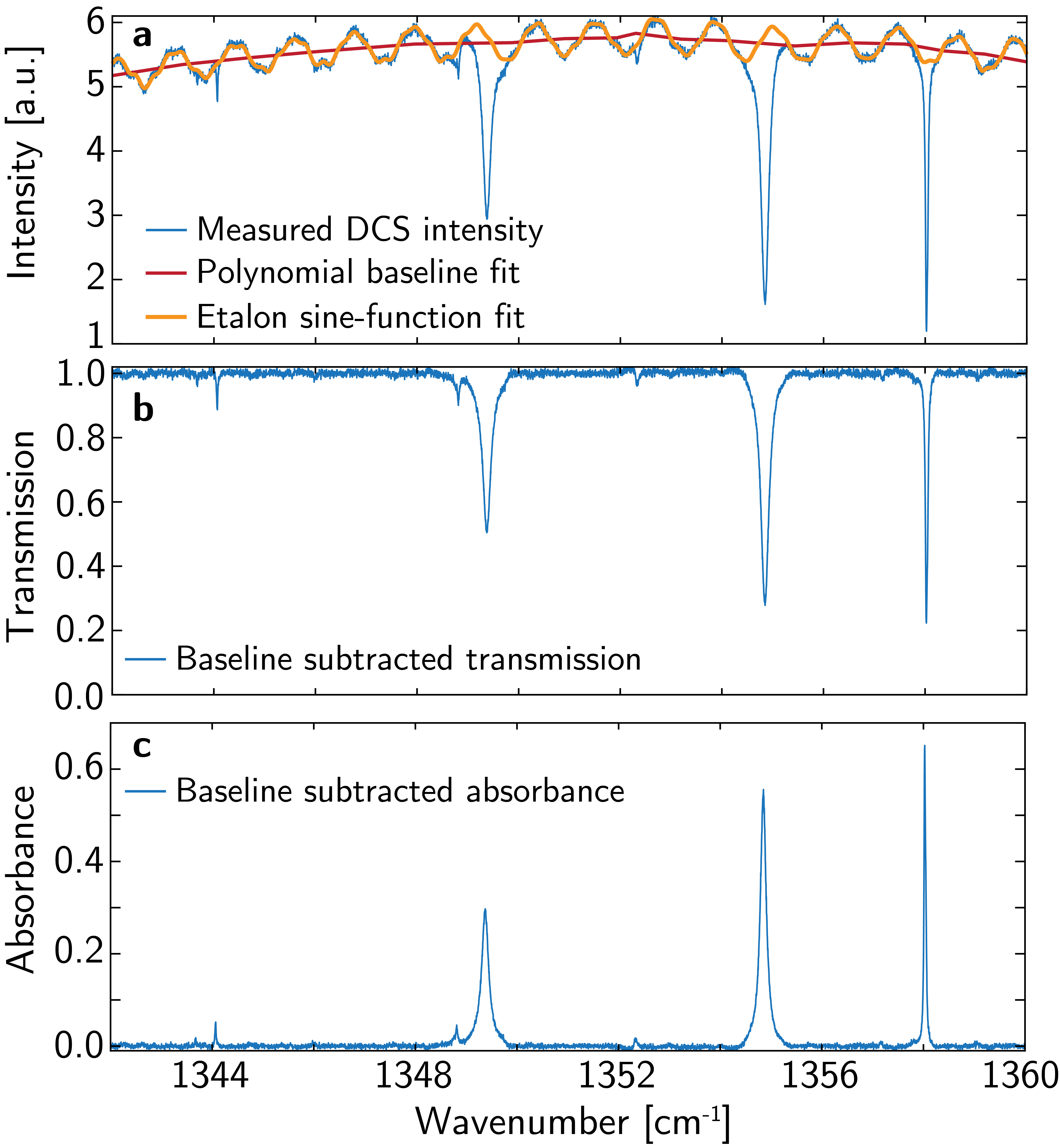}
\caption{\label{FigWater}Baseline correction procedure for atmospheric water absorption. (a) The measured transmitted spectra of the dual-comb system (blue curve) along with the fitted low-order polynomial baseline (red) and the fitted sine-function (orange) to correct for the mid-IR etalons. (b) The baseline subtracted transmission spectra along with the (c) absorbance spectra of the atmospheric water. Presented data is measured with the waveguide width of 2.95~\textmu{}m and plotted as an example to clarify our baseline correction procedure for all other frequency ranges reported in the manuscript.}
\end{figure}

To calculate the absorbance of the atmospheric water, we first need to remove the spectral baseline and mid-IR etalons from the measured dual-comb transmitted spectrum (Fig.~\ref{FigWater}a). We estimate the spectral baseline using a low-order polynomial fit (red curve in Fig.~\ref{FigWater}a) with added sine-function terms to account for the mid-IR etalons (orange curve in Fig.~\ref{FigWater}a). We then calculate the transmission spectrum of the dual-comb system (Fig.~\ref{FigWater}b) as $I/I_{0}$, where $I_{0}$ and $I$ are the calculated baseline and the measured transmitted intensity spectrum, respectively. As the last step the absorbance (Fig.~\ref{FigWater}c) is defined according to the Beer-Lambert law as $A = \mathrm{log_{10}}(I/I_{0})$. 

For comparison to HITRAN, we use the water absorption cross-sections and line intensities from the database in a Voigt fitting algorithm. This algorithm accounts for temperature dependent Doppler shift of the center frequencies, Doppler broadening with Gaussian line shape, and Lorentzian line shape pressure dependent broadening. We estimate the atmospheric pressure and water concentration by fitting the Voigt line shapes and intensities to the DCS data. In doing so, we keep the experimental temperature constant at 296~K while beam path length, experimental atmospheric pressure and volume-mixing ratio of water content are fitted into the algorithm to minimize the residuals between DCS data and HITRAN prediction. The fitting is performed on multiple spectral peaks within all frequency ranges accessed by the three waveguide widths of 3.35~\textmu{}m, 2.95~\textmu{}m, and 2.80~\textmu{}m. We then report the mean value of all fitted atmospheric pressures and volume-mixing ratios of the water content as the environmental parameters of the experiment with the RMS difference between individual values and the calculated mean as the error bar of the estimated values.

\begin{figure*}[t!]
\includegraphics[width=\textwidth]{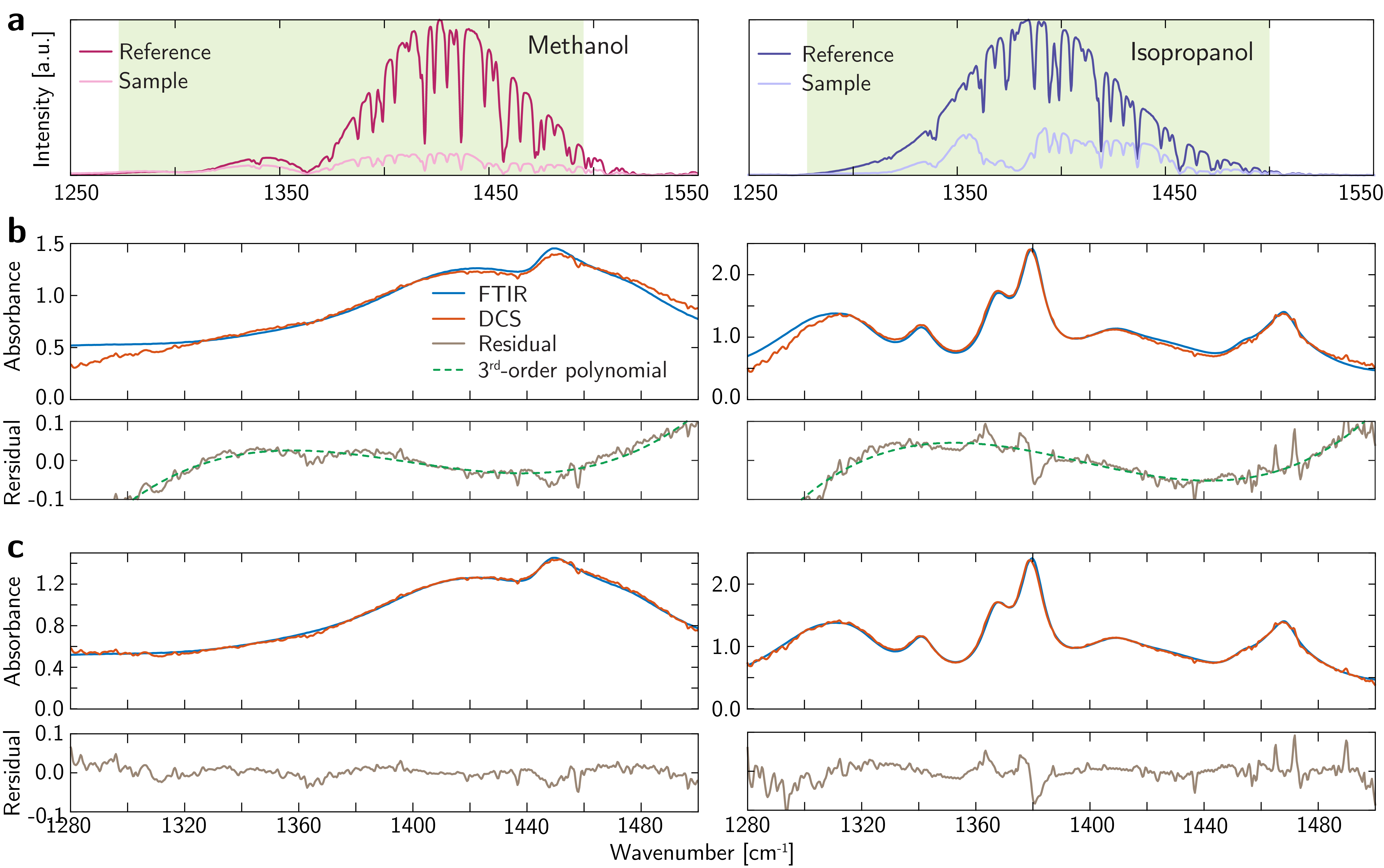}
\caption{\label{FigBASE}Broadband baseline correction procedure for methanol and isopropanol dual-comb data. (a) The simultaneously measured reference and sample spectra for methanol and isopropanol on left and right columns, respectively. (b) The DCS measured absorbance spectra of isopropanol and methanol compared with FTIR measurements. The sample data is normalized to the reference for baseline subtraction. The experimental residuals are calculated as DCS data minus FTIR and presented in the bottom panels along with the 3\textsuperscript{rd}-order polynomial fits to the residuals. The polynomial fits to both methanol and isopropanol residuals result in very similar polynomial curves. This similarity is a signature of a systematic different between our dual-comb system and the commercial FTIR used as the reference. (c) Comparison between DCS data and FTIR after removing the fitted 3\textsuperscript{rd}-order polynomial function from the DCS data. The new residuals are calculated as residuals in (b) minus the polynomial fit. The polynomial fitting acts as a second degree baseline correction mechanism.}
\end{figure*}

\section{Broadband baseline correction}
 
The dual-comb measured sample and reference spectra are presented in Fig.~\ref{FigBASE}a for methanol and isopropanol on left and right columns, respectively. To calculate the absorbance from the Beer-Lambert law, we initially remove the spectral baseline by dividing the sample spectra to their respective reference measurement. We compare the results with identical measurements performed using a commercial Fourier transform infrared spectrometer (FTIR) in Fig.~\ref{FigBASE}b. The dual-comb spectroscopy (DCS) data agrees well with the FTIR, however, the residuals for both chemicals show a very similar 3\textsuperscript{rd}-order polynomial pattern. This pattern is studied further by fitting individual polynomial curves to both residuals. We realize that the same polynomial curve fits to both data. The fitted function is defined as $R(f) = a_{3}f^{3}+a_{2}f^{2}+a_{1}f+a_{0}$, where $R$ is the fitted polynomial function and $f$ is the frequency. $a_{3} = \num{2.49e-7}$, $a_{2} = \num{-0.001}$, $a_{1} = \num{1.46}$, and $a_{0} = \num{-678}$ are the fitted 3\textsuperscript{rd}-order polynomial coefficients.

The fact that the same 3\textsuperscript{rd}-order polynomial function explains the residual baselines of two independent measurements, points to a systematic difference between our dual-comb system and the commercial FTIR used in this experiment. We account for the observed different between DCS and FTIR measurements by removing the fitted polynomial function from the DCS data. Results of this step are presented in Fig.~\ref{FigBASE}c. There is an excellent agreement between the DCS and FTIR data after the 3\textsuperscript{rd}-order polynomial baseline correction with spectral residuals limited to the uncorrected atmospheric water absorption in this spectral region.

\section{absorbance peak assignments}

\begin{figure}[t!]
\includegraphics[width=\linewidth]{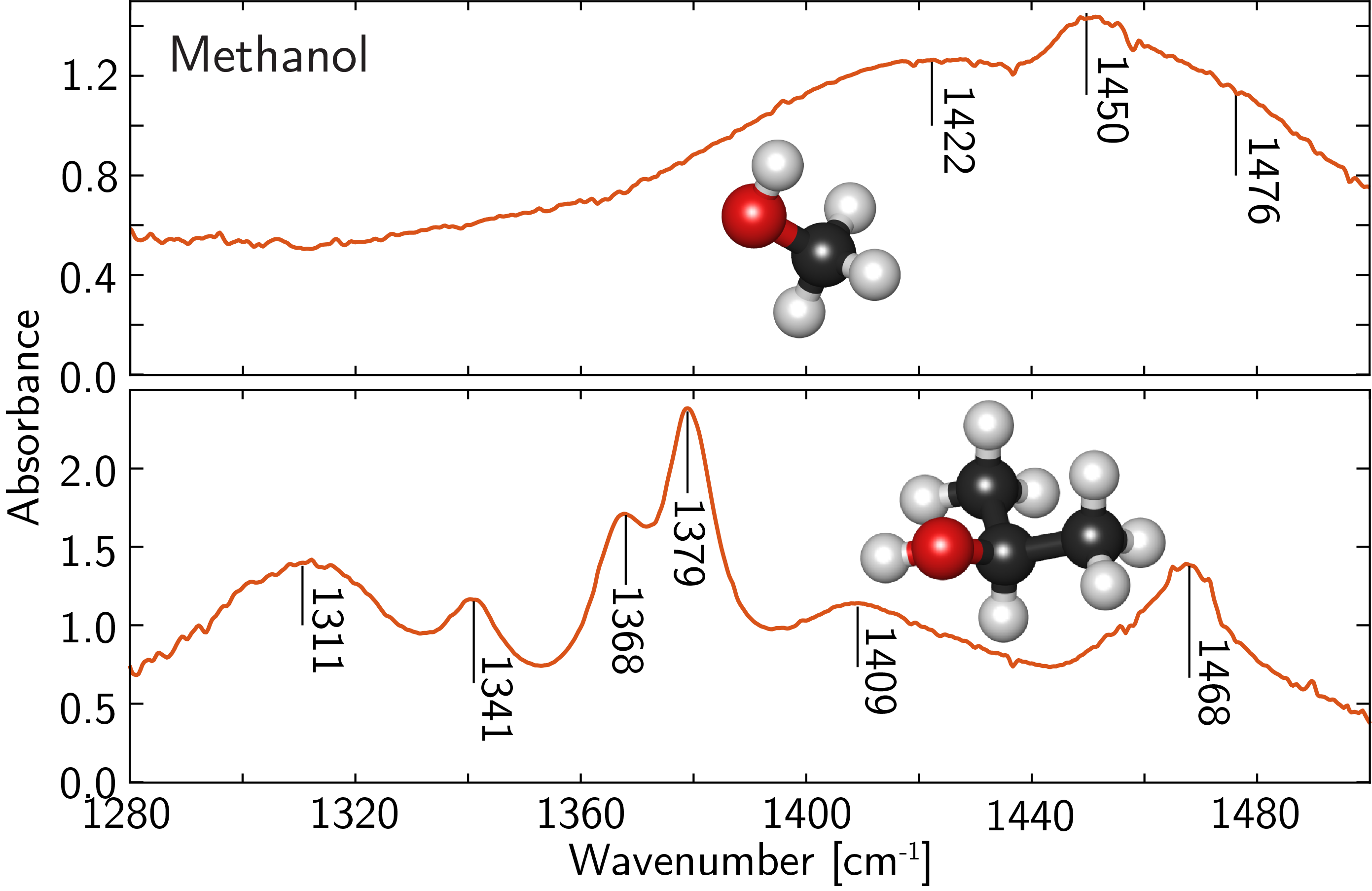}
\caption{\label{FigDCS}The DCS absorbance spectra of methanol and isopropanol, with the frequency of the measured O---H and C---H bending vibration lines marked.}
\end{figure}

We identify the measured absorbance peaks of methanol and isopropanol as various C---H and O---H bending vibrations. The most compelling infrared signature is the absorbance peak doublet around 1360--1400~\si{\per\cm} in the isopropanol spectrum (Fig.~\ref{FigDCS}). This is a characteristic of the presence of a geminal dimethyl group, two CH\textsubscript{3} groups attached to a central carbon atom carrying an OH group (\cite{klein_organic_2012,perkin}, insets of Fig.~\ref{FigBASE}). This is corroborated by the weaker lower frequency band indicative of a secondary alcohol \cite{perkin}. Such observation in this region can be used to non-destructively identify the presence of secondary and tertiary alcohols with specificity that the 2--3 \textmu{}m spectral region does not offer \cite{klein_organic_2012}.

For a more specific assignment of the measured peaks, we use the Gaussian 03 and GaussView computational chemistry softwares \cite{gaussView} to simulate the relative peak positions and integrated intensities of the different C---H and O---H bending vibrations in liquid phase methanol and isopropanol. For methanol, three spectral features are recorded, two spectral peaks at 1422~\si{\per\cm} and 1450~\si{\per\cm} with a spectral shoulder at 1476~\si{\per\cm}. Reference \cite{Herzberg} identifies these as $\nu\textsuperscript{$\prime$}\textsubscript{6}$\textsuperscript{CH\textsubscript{3}}, $\nu\textsuperscript{$\prime\prime$}\textsubscript{6}$\textsuperscript{CH\textsubscript{3}}, and $\nu\textsubscript{3}$\textsuperscript{CH\textsubscript{3}} bending vibrations, respectively. We assign the 1420~\si{\per\cm} peak to CH\textsubscript{3} bending with an umbrella-like wagging motion of hydrogen atoms. The 1450~\si{\per\cm} peak is assigned as the CH\textsubscript{3} bending with a scissoring motion in all three hydrogen atoms. The spectral shoulder at 1476~\si{\per\cm} is assigned to the CH\textsubscript{3} bending vibration with two of the hydrogen atoms undergoing scissoring motion. 

For isopropanol, the following six absorbance peaks are measured (Fig.~\ref{FigDCS}b): (1) O---H bending in symmetry-plane of the molecule at 1311~\si{\per\cm}, (2) wagging motion of the hydrogen atom in O---C---H group at 1341~\si{\per\cm}, (3) O---C---H group hydrogen bending in symmetry-plane of the molecule at 1368~\si{\per\cm}, (4) asymmetrical wagging motion in the CH\textsubscript{3} group at 1379~\si{\per\cm}, (5) symmetrical umbrella like wagging motion in the CH\textsubscript{3} along with in-plane bending of the O---H group at 1409~\si{\per\cm}, and (6) scissoring motion in the CH\textsubscript{3} while the O---H group is stationary at 1468~\si{\per\cm}. It is worth noting that Gaussian 03 calculations and GaussView visualizations are of the isolated molecule's eigenfrequencies and eigenmodes. The measured frequencies in the condensed phase samples are shifted due to interactions in the liquid. Assignments were made based on the eigenfrequencies and reported motions from GaussView \cite{gaussView} and Herzberg \cite{Herzberg}.

\bibliography{References}